\documentclass[journal,12pt, draftclsnofoot, onecolumn]{IEEEtran}

\usepackage[cmex10]{amsmath}
\usepackage{amssymb}
\usepackage{amsthm}
\usepackage{graphicx}
\usepackage{array,color}
\usepackage{amsmath}
\usepackage{stfloats}
\usepackage{graphicx}
\usepackage{subfigure}
\usepackage{tabularx}
\usepackage{epsfig,epsf,color,balance,cite}
\usepackage{algorithmic}
\usepackage{algorithm}
\usepackage{bm}
\usepackage{textcomp}
\usepackage{caption}
\usepackage{multirow}
\usepackage{cite}
\usepackage{enumerate}
\usepackage{cases} 
\usepackage{color}
\usepackage{url}
\renewcommand{\baselinestretch}{1.40}

\begin{document}
\title{Resource Allocation for D2D Communications Underlaying a NOMA-Based Cellular Network}
\author{Yijin Pan, Cunhua Pan, Zhaohui Yang, Ming Chen \\
\thanks{Y. Pan, Z. Yang and M. Chen are with the National Mobile Communications Research Laboratory, Southeast University, Nanjing 211111, China. Email:\{panyijin,yangzhaohui,chenming\}@seu.edu.cn.}
\thanks{C. Pan is with the School of Electronic Engineering and Computer Science, Queen Mary, University of London, London E1 4NS, UK. Email: c.pan@qmul.ac.uk.}
\vspace{-2.5em}}
\maketitle

\vspace{-1.0em}
\begin{abstract}
\vspace{-1.0em}
This letter investigates the power control and channel assignment problem in device-to-device (D2D) communications underlaying a non-orthogonal multiple access (NOMA) cellular network.
With the successive interference cancellation decoding order constraints, our target is to maximize the sum rate of D2D pairs while guaranteeing the minimum rate requirements of NOMA-based cellular users.
Specifically, the optimal conditions for power control of cellular users on each subchannel are derived first. 
Then, based on these results, we propose a dual-based iterative algorithm to solve the resource allocation problem.
Simulation results validate the superiority of proposed resource allocation algorithm over the existing orthogonal multiple access scheme.
\end{abstract}


\vspace{-1em}
\section{Introduction}
The device-to-device (D2D) communications have been considered as a promising way to alleviate the upcoming traffic pressure on core networks.
Due to the short transmission distance of D2D pairs, the spectrum efficiency can be significantly improved by the spectrum reuse with cellular users (CUs). 
In conventional networks, the uplink resources are normally provided for D2D communications since traffic in downlink is significantly heavier than that in uplink \cite{EquDang,GraphD2D}. 
However, some uplink applications with high rate requirements are becoming popular in the future networks such as the Skype video call\cite{ULDL}. 
Hence, the traffic between uplink and downlink is becoming less asymmetric and the resource allocation problem for D2D communications underlaying the downlink celluar network should be studied as well \cite{SPL,yangzhaohui,MalandrinoDownlink}.

Apart from D2D communications, non-orthogonal multiple access (NOMA) is another emerging technology to handle the transmission pressure in the near future \cite{NOMAMag}.
In a NOMA-based cellular network, multiple CUs are allowed to share the same subchannel via different power levels, and successive interference cancellation (SIC) is adopted at the CUs for decoding.
In this way, the NOMA-based cellular network can greatly increase system throughput and allow massive connectivities.
Recently, several approaches have been proposed to combine the D2D communications with NOMA technology \cite{JingZhao,7542601}.  
The D2D users were grouped through the NOMA way in \cite{JingZhao} to achieve better D2D rate performance, and the channel allocation problem for the NOMA-based D2D groups is modeled as a Many-to-One matching.
Furthermore, the D2D assisted NOMA scheme was proposed in \cite{7542601} to enhance system throughput performance. 
D2D pairs were merely assumed to transmit on exclusive channels without sharing channels with CUs in \cite{7542601}, even though frequency reuse between D2D pairs and CUs is spectrum efficient. 
However, when D2D pairs reuse spectrum with NOMA-based CUs, the co-channel interference in SIC decoding will become further complicated, which may destroy the original SIC decoding order of CUs.
To conquer this issue, one should impose an additional restriction for the power control and channel assignment of the D2D pairs, which has not been studied in current literature.
This motivates us to reconsider the resource allocation problem for D2D communications when the D2D pairs share spectrum with the NOMA-based CUs. 

In this letter, we consider the power control and channel assignment for the D2D pairs underlaying NOMA-based cellular networks with consideration of the SIC decoding constraints.
This scenario is different from that in \cite{JingZhao}, where NOMA is implemented in D2D transmissions. Furthermore, since power control is not considered for either CUs or NOMA-based D2D groups in \cite{JingZhao}, the approach in \cite{JingZhao} cannot be directly applied to solve our problem.
Our target is to maximize the sum rate of D2D pairs while guaranteeing the minimum rate requirements of CUs.
We derive the optimal conditions for power control of the NOMA-based CUs first, then propose a dual-based iterative algorithm to solve the resource allocation problem.
Specifically, by adopting the auxiliary variables and relaxing the binary constraints, the formulated optimization problem is transformed into a convex one, which can be optimally solved by the dual method.
Finally, simulation results show the significant D2D sum rate gains of proposed algorithm over the conventional orthogonal multiple access (OMA) scheme.

\vspace{1.25em}
\section{System Model and Problem Formulation}

Consider a downlink NOMA-based cellular network, where base station (BS) serves CUs through $N$ subchannels (SCs). 
By adopting NOMA, $M$ CUs are multiplexed in the same SC by splitting them in the power domain\footnote{It is assumed that CUs served by the same SC are already scheduled. The CU grouping issues (see e.g. \cite{NOMAMag}) are generally the task of the upper-layer, which are beyond the scope of this paper.}.
That is to say, the total number of CUs is $NM$.
Meanwhile, there are $K (K \leq N)$ underlaid D2D pairs randomly distributed in the cell.

Denote $\mathcal K = \{ 1,\cdots,K\}$ and $\mathcal{N}=\{ 1,\cdots,N\}$ as the sets of D2D pairs and SCs, respectively.
The superposition symbol transmitted by BS on SC $n$ to CUs is 
\begin{equation}
x^n = \sum_{i=1}^{M}\sqrt{p^n_i}s^n_i,
\end{equation}
where $s^n_i$ and $p^n_i$ are the transmit signal and transmit power for CU $i$ on SC $n$, respectively.

To implement NOMA, BS needs to inform each CU of the SIC decoding order, so that strong CUs can decode and remove the signal from weak CUs.
In current work, it is generally assumed that the SIC decoding order follows the increasing order of channel gains \cite{7890454,Ding2014}.
Let $h^n_{i}$ denote the channel from BS to CU $i$ on $n$-th SC\footnote{Similar as \cite{7890454,Ding2014}, the receivers are assumed to have the perfect channel state information by channel feedback.}.
When $|h^n_{1}| \!\leq |h^n_{2}| \!\leq\!\!\cdots\!\!\leq \!|h^n_{M}|$, CU $i$ can successfully decode and remove the interference from CU $j, \forall j < i$.
However, in our work, underlaid D2D pairs also contribute to the co-channel interference, which affects the NOMA decoding order.

In this case, the received SINR at CU $i$ to decode the signal $s_j, j < i$, on SC $n$ is 
\begin{equation}\label{SINR}
\text{SINR}^{n}_{i \to j} = \frac{p^n_j|h^n_i|^2}{|h^n_i|^2\sum_{t=j+1}^{M} p^n_t+\sum_{k=1}^K \alpha^n_k q^n_k |h^n_{k,i}|^2+ \sigma^2},
\end{equation}
where the binary variable $\alpha^n_k$ denotes whether or not SC $n$ is assigned to D2D pair $k$.
The term $\sum_{k=1}^K \alpha^n_k q^n_k |h^n_{k,i}|^2$ represents the co-channel interference to CU $i$ from the underlaid D2D pairs on SC $n$,
where $q^n_{k}$ is the transmit power of D2D pair $k$ and $|h^n_{k,i}|$ represents the channel gain from D2D pair $k$ to CU $i$ on SC $n$.
Specifically, when CU $i$ desires to decode the signal of CU $j$ from superposition symbol $x^n$, the interference cancellation is successful if the CU $i$'s received SINR is larger or equal to CU $j$'s own received SINR. 
Therefore, to protect the given SIC decoding order, the following conditions should be satisfied. 
\begin{equation}
\frac{\sum_{k=1}^K \alpha^n_k q^n_k  |h^{n}_{k,j}|^2 + \sigma^2}{|h^n_j|^2}\geq \frac{\sum_{k=1}^K \alpha^n_k q^n_k |h^{n}_{k,{i}}|^2 + \sigma^2}{|h^n_{i}|^2}, \label{cx}
\end{equation}
\vspace{-0.3em}for $i,j \in \{1,\cdots,M\} \triangleq \mathcal{M}$, $ j < i$, and $n \in \mathcal{N}$.
The set $\mathcal{M}$ represents the set of CUs' index on each SC. 
Note that there will be $\frac{M(M-1)}{2}$ constraints for each SC in the form of (\ref{cx}). 
To simplify the decoding order constraints, the following equivalent inequalities can be implied from (\ref{cx}) 
\begin{equation}
\frac{\sum_{k=1}^K \alpha^n_k q^n_k  |h^{n}_{k,i}|^2 + \sigma^2}{|h^n_i|^2}\geq \frac{\sum_{k=1}^K \alpha^n_k q^n_k |h^{n}_{k,{i+1}}|^2 + \sigma^2}{|h^n_{i+1}|^2}, \label{c0}
\end{equation}
for $i \in \mathcal{M}\setminus \{M\}$, and $n \in \mathcal{N}$.
In this form, there are only $M-1$ constraints on each SC.

The achievable rate of CU $i$ on SC $n$ in bits/s/Hz is 
\begin{equation}\label{Rij}
R^{n}_{i \to i}= \log_2(1+\text{SINR}^{n}_{i \to i}).
\vspace{-0.75em}
\end{equation}
Although the spectrum efficiency can be improved by allowing multiple D2D pairs reusing the same SC, the design of efficient resource allocation schemes in this paradigm requires high computation complexity.
Moreover, heavy signaling overhead exchange occurs since channel state information of the interference channels between different D2D pairs over all SCs should be estimated. 
Given theses concerns, we assume that one SC is only allocated to at most one D2D pair.
Consequently, we have the following channel assignment constraints:
\begin{eqnarray}
\sum_{k=1}^K  \alpha^{n}_{k} \leq 1,  \alpha^{n}_{k} \in \{0,1\}, \forall n \in \mathcal{N}, k \in \mathcal{K}. \label{c1}
\end{eqnarray}
The SINR at the receiver of D2D pair $k$ on SC $n$ is 
\begin{equation} \label{S}
\text{SINR}^n_{k} = \frac{q^n_{k}|g^n_{k}|^2}{ |g^{n}_{k,B}|^2\sum_{i=1}^{M} p^n_{i} + \sigma^2},
\end{equation}
where $|g^n_{k}|$ is the channel gain between the transmitter and receiver of D2D pair $k$ on SC $n$, and $|g^{n}_{k,B}|$ is the interference channel gain from BS to the receiver of D2D pair $k$ on SC $n$.
In this case, the achievable rate of D2D pair $k$ on SC $n$ in bits/s/Hz is
\vspace{-0.5em}
\begin{equation} \label{SNR}
R^n_{k}= \log_2(1+\text{SINR}^n_{k}).
\end{equation}
Meanwhile, to guarantee the rate fairness among CUs, the minimum rate requirements for CUs are imposed as
\vspace{-1.0em}
\begin{equation}
R^{n}_{i \to i} \geq \gamma_i^n, \forall i \in \mathcal{M}, n \in \mathcal{N}, \label{c2}
\end{equation}
where $\gamma_i^n$ is the rate requirement of CU $i$ on SC $n$.
The transmit power constraints for D2D pairs and BS for CUs are 
\vspace{-1.0em}
\begin{equation}
\sum_{n=1}^N \alpha_k^n q^n_{k} \leq P^D_{\max} , \forall k \in \mathcal{K}, \label{c3}
\end{equation}
\vspace{-1.0em}
\begin{equation}
\sum_{i=1}^Mp^n_{i} \leq P^C_{\max}, \forall n \in \mathcal{N}.\label{c4}
\end{equation}

To maximize the sum rate of D2D pairs, the following optimization problem is obtained.
\begin{subequations}
\begin{align}
\mathcal{P}1:\mathop{\max }_{\{p^n_{k},\alpha^n_k, q^n_{k}\}}
                   &\quad  R_{\text{max}}^D=\sum_{k =1}^{K}  \sum^N_{n=1}  \alpha^n_{k} R^n_{k} ,    \label{opt1}   \\
\textrm{s.t.} &\quad (\ref{c0}), (\ref{c1}), (\ref{c2})-(\ref{c4}). \nonumber        
\end{align}
\end{subequations}

\section{Power Control and Channel Assignment}
To solve the sum rate maximization Problem $\mathcal{P}1$, the optimal conditions for power control of the NOMA-based CUs on a given SC are first investigated. 
Then, we propose a dual-based iterative method to obtain the power control and channel assignment for D2D pairs.

\subsection{Optimal Power Control for CUs}

If SC $n$ is assigned to D2D pair $k$, we first determine the optimal transmit power conditions for CUs.
For simplicity, the superscript $n$ is omitted in the following analysis of this subsection.
To solve the power control problem, we define 
\begin{equation}
\xi_{k,i}=\frac{|h_{k,i}|^2}{|h_i|^2} ,   \Delta_i=\frac{\sigma^2}{|h_i|^2} , \forall i \in \cal M. 
\end{equation}
It is easy to know that the constraint (\ref{c2}) should hold with equality for the optimal transmit power of CU $i$ denoted as $p_{i}^{\ast}, \forall i \in \mathcal{M}$\cite{SPL,yangzhaohui}.
Otherwise, the sum rate of D2D pairs can be further improved by decreasing $p_{i}^{\ast}$.
Setting $(\ref{c2})$ with equality for CU $M$, we have 
\begin{equation}\label{pm}
p_M^{\ast} = (2^{\gamma_M} - 1) \left( q_k \xi_{k,M} + \Delta_{M} \right).
\end{equation}
Accordingly, for the CU $i, \forall i \in \mathcal{M}\setminus \{M\}$,  the optimal transmit power $p_{i}^{\ast}$ is 
\begin{equation}\label{Dt}
p_{i} ^{\ast}= \left(2^{\gamma_i}-1\right)\left(q_k \xi_{k,i} + \Delta_{i} + \sum_{t=i+1}^M p_t^{\ast} \right).
\end{equation}
It is not easy to obtain the explicit expression of $p_{i}^{\ast}$ from (\ref{Dt}). 
To assist solving this issue, we define $S_i=\sum_{t=i}^M p_{t}^{\ast}$, 
which represents the summation of transmit powers from CU $i$ to CU $M$.
Substituting $S_i$ in (\ref{Dt}), we have 
\begin{equation}
S_i = (2^{\gamma_i} -1) (q_k \xi_{k,i} + \Delta_i)+ 2^{\gamma_i} S_{i+1}.
\end{equation}
According to the recursive relations, we can obtain that 
\begin{equation} \label{mid}
S_i\!\!=\!\!\!\!\sum_{j=0}^{M-i-1}2^{\sum_{l=0}^{j-1} \gamma_{i+l}}(2^{\gamma_{i+j}}-1)(q_k \xi_{k,i+j} \!+ \!\Delta_{i+j})+2^{\sum_{s=0}^{M-i-1} \gamma_{i+s}} S_M.
\end{equation}
where we define $2^{\sum_{l=0}^{-1} \gamma_{i+l}}=2^0=1$.
Obviously, $S_M= p_M^{\ast}$.
Based on (\ref{pm}), (\ref{mid}) is simplified to 
\begin{equation}\label{sum}
\!\!S_i = \sum_{j=0}^{M-i}2^{\sum_{l=0}^{j-1} \gamma_{i+l}}(2^{\gamma_{i+j}}-1)(q_k \xi_{k,i+j} \!+ \!\Delta_{i+j}).
\end{equation}
In addition, the optimal transmit power for CU $i, \forall i \in \mathcal{M}\setminus \{M\} $ is obtained by $p_{i}^{\ast} = \sum_{t=i}^M p_{t}^{\ast}-\sum_{t=i+1}^M p_t^{\ast}= S_i -S_{i+1}$.
By further using (\ref{sum}) and defining $\sum_{l=1}^{0} \gamma_{i+l} =2^0=1$, we have
\begin{equation}\label{pi}
p_{i}^{\ast} = (2^{\gamma_{i}}-1) \sum_{j=1}^{M-i}2^{\sum_{l=1}^{j-1} \gamma_{i+l}} (2^{\gamma_{i+j}}-1)(q_k \xi_{k,{i+j}} + \Delta_{i+j})+  (2^{\gamma_{i}}-1)(q_k \xi_{k,{i}} + \Delta_{i}), i \in \mathcal{M}\setminus \{M\}.  
\end{equation}

\subsection{D2D Power Control and Channel Assignment}
Based on the previous analysis, the transmit powers for CUs are all determined by the transmit power of the multiplexed D2D pair.
When SC $n$ is assigned to D2D pair $k$, according to (\ref{sum}), the transmit power constraint (\ref{c4}) is equivalent to
\begin{equation} \label{c7}
\sum_{j=0}^{M-1}\Gamma_j (q^n_k \xi^n_{k,1+j} \!+ \!\Delta^n_{1+j}) \leq P^C_{\max}.
\end{equation}
where $\Gamma_j = 2^{\sum_{l=1}^{j} \gamma_l}(2^{\gamma_{1+j}}-1)$.
Then, the following inequality holds.
\begin{equation}
 q_k^n\leq \frac{P^C_{\max}-\sum_{j=0}^{M-1}\Gamma_j\Delta^n_{1+j}}{\sum_{j=0}^{M-1}\Gamma_j\xi_{k,1+j}^n}. 
\end{equation}
According to (\ref{sum}), the SIC successful decoding order constraints in (\ref{c0}) can be rewritten as
\begin{equation} \label{c6}
 q^n_k \xi^n_{k,i} + \Delta^n_i \geq q^n_k \xi^n_{k,i +1}+ \Delta^n_{i+1}, \forall i \in \mathcal{M}\setminus \{M\}. 
\end{equation}
Recall that $\Delta^n_{i+1} \leq \Delta^n_{i}$ since $|h^n_{i}| \leq |h^n_{i+1}|$.
Note that if $\xi^n_{k,i} \geq \xi^n_{k,i+1}$, (\ref{c6}) is feasible for any non-negative $q^n_k$. 
Hence, the transmit power of D2D pair $k$ on SC $n$ should satisfy 
\begin{equation}\label{pk}
q^n_k \leq  \min\limits_{\{ i \in \mathcal{M}\setminus \{M\} |\xi^n_{k,i}< \xi^n_{k,i +1}\}} \left\{ \frac{\Delta^n_{i+1}-\Delta^n_{i}}{\xi^n_{k,i}-\xi^n_{k,i +1}}\right\}.
\end{equation}
\textit{Remark}: 
			According to (\ref{pk}), we find that if $\xi^n_{k,i}< \xi^n_{k,i +1}$, i.e., $\frac{|h^n_{k,i}|^2}{|h^n_i|^2} < \frac{|h^n_{k,i+1}|^2}{|h^n_{i+1}|^2}$, one additional transmit power constraint is imposed on the D2D pair to protect the SIC decoding order of CUs.
			On the other hand, if the condition $\xi^n_{k,i} \geq \xi^n_{k,i +1}$, i.e., $\frac{|h^n_{k,i}|^2}{|h^n_i|^2}\geq \frac{|h^n_{k,i+1}|^2}{|h^n_{i+1}|^2}$ holds for all $ i \in \mathcal{M}\setminus \{M\}$, the SIC decoding order constraints in (\ref{c0}) are alway satisfied.

Substituting (\ref{sum}) into (\ref{SNR}), the achievable rate of D2D pair $k$ on SC $n$ is
\begin{equation}\label{Rkn}
R^n_{k}(q^n_{k})=\log_2\left(1+ \frac{d^n_k q^n_{k}}{ q^n_k+e^n_k} \right),
\end{equation}
where $d^n_k=\frac{|g^n_{k}|^2}{|g^n_{k,B}|^2\sum_{j=0}^{M-1} \Gamma_j \xi_{j+1}}$, $e^n_k = \frac{|g^n_{k,B}|^2\sum_{j=0}^{M-1} \Gamma_j \Delta_{j+1} + \sigma^2}{|g^n_{k,B}|^2\sum_{j=0}^{M-1} \Gamma_j \xi_{j+1}}$.
Given the above results, the original Problem $\mathcal{P}1$ is simplified to
\begin{subequations}
\begin{align}  
\mathcal{P}2: \mathop{\max }_{\{\alpha^n_k,q^n_{k}\}} & \quad R_{\text{max}}^D =\sum_{k =1}^K \sum_{n=1}^N \alpha_k^n R^n_{k}(q^n_{k}),  \label{Rmax} \\
\textrm{s.t.} &\quad  (\ref{c1}), (\ref{c3}), \\
&\quad  0 \leq q^n_k \leq Q^n_k,  \forall k \in \mathcal{K},n \in \mathcal{N}, \label{st_Qkn}
\end{align}
\end{subequations}
where
$
Q^n_k = \min\left\{\max\left\{0, \frac{P^C_{\max}-\sum_{j=0}^{M-1}\Gamma_j\Delta^n_{1+j}}{\sum_{j=0}^{M-1}\Gamma_j\xi_{k,1+j}^n}\right\}, \min\limits_{\{ i \in \mathcal{M}\setminus \{M\} |\xi^n_{k,i}< \xi^n_{k,i +1}\}} \left\{ \frac{\Delta^n_{i+1}-\Delta^n_{i}}{\xi^n_{k,i}-\xi^n_{k,i +1}}\right\}\right\}.
$

It is easy to see that $f(q^n_k)=\frac{d^n_k q^n_{k}}{ q^n_k+e^n_k}$ is concave with respect to (w.r.t) $q^n_{k}$.
Consequently, $R^n_{k}(q^n_{k})$ is concave w.r.t $q^n_{k}$ due to that the logarithmic function is increasing and concave \cite[Page 84]{boyd2004convex}.
However, problem $\mathcal{P}2$ is not convex due to (\ref{Rmax}) and (\ref{c1}).
By introducing the auxiliary variable $x^n_k= \alpha^n_k q^n_{k}$, and temporarily relaxing the integer constraints for $\{\alpha^n_k\}$, Problem $\mathcal{P}2$ is transformed into
\vspace{-1.5em}
\begin{subequations}
	\begin{align}  
	\mathcal{P}3: \underset{ \underset{x^n_{k} \in [0,\alpha_k^nQ^n_k]}{\alpha^n_k \in [0,1]}}{\mathop{\max }} &\quad R_{\text{max}}^D(\alpha_k^n,x^n_k) =\sum_{k =1}^K \sum_{n=1}^N \alpha_k^n R^n_{k}\left(\frac{x^n_k}{\alpha_k^n}\right), \label{P3obj} \\
	\textrm{s.t.} &\quad  \sum_{n=1}^N x^n_{k} \leq P^D_{\max} , \forall k \in \mathcal{K}, \label{st_pw} \\
	&\quad \sum_{k=1}^K  \alpha^{n}_{k} \leq 1, \forall n \in \mathcal{N}.  \label{st_asg}
	\end{align}
\end{subequations}
It is inferred that $R_{\text{max}}^D(\alpha_k^n, x^n_k)$ is concave w.r.t $(\alpha_k^n,x^n_k)$ within a triangular region due to the perspective property \cite{SPL,99Jsac}, so that Problem $\mathcal{P}3$ is convex.
Therefore, the optimal solution to Problem  $\mathcal{P}3$ can be obtained by using the standard dual method.
The Lagrangian is obtained as
\begin{equation}
\mathcal{L} =  \sum_{k =1}^K \sum_{n=1}^N \alpha_k^n R^n_{k}\left(\frac{x^n_k}{\alpha_k^n}\right) +\sum_{k =1}^K \lambda_k\left(P^D_{\max} - \sum_{n=1}^N x_k^n\right) + \sum_{n=1}^N \beta_n\left(1-\sum_{k=1}^K  \alpha^{n}_{k}\right),
\end{equation}
where $\{\lambda_k\}$ and $\{\beta_n\}$ are the non-negative dual variables associated with the constraints (\ref{st_pw}) and (\ref{st_asg}), respectively.
Taking the derivative of $\mathcal{L}$ w.r.t $x_k^n$ and $\alpha^{n}_{k}$ respectively, we have
\begin{eqnarray}
\frac{\partial \mathcal{L}}{\partial x_k^n} =  {R^n_{k}}'\left(\frac{x^n_k}{\alpha_k^n}\right) - \lambda_k, \quad
\frac{\partial \mathcal{L}}{\partial \alpha^{n}_{k}} = {R^n_{k}}\left(\frac{x^n_k}{\alpha_k^n}\right) - \frac{x^n_k}{\alpha_k^n} {R^n_{k}}'\left(\frac{x^n_k}{\alpha_k^n}\right) -\beta_n,
\end{eqnarray}
where ${R^n_{k}}'(t)$ is the derivative of $R^n_{k}(t)$ w.r.t $t$.
Applying the Karush-Kuhn-Tucker conditions, we can obtain the following necessary conditions for the optimal solution $({\alpha_k^n}^*, {x_k^n}^*)$. 

If ${\alpha_k^n}^* =0$, then ${x_k^n}^* =0$, and we have $\frac{\partial \mathcal{L}}{\partial x_k^n} <0, \frac{\partial \mathcal{L}}{\partial \alpha_k^n} <0$,
for all $\alpha_k^n \in (0,1]$ and $x_k^n \in (0,Q_k^n]$.

If  ${\alpha_k^n}^* \neq 0$, we have 
\begin{equation} \label{div2}
\frac{\partial \mathcal{L}}{\partial x_k^n} 
 \left\{\begin{array}{llll}
< 0,& \text{if } {x_k^n}^* =0  \\ 
=0, & \text{if } {x_k^n}^* \in (0, Q_k^n)  \\ 
>0, & \text{if }  {x_k^n}^* =Q_k^n
\end{array}\right 
., \quad \frac{\partial \mathcal{L}}{\partial \alpha_k^n} 
\left\{\begin{array}{llll}
=0, & \text{if } {\alpha_k^n}^* \in (0, Q_k^n)  \\ 
>0, & \text{if }  {\alpha_k^n}^* =1.
\end{array}\right.
\end{equation}

When ${x_k^n}^* \in (0, Q_k^n)$, ${x_k^n}^*$ can be obtained by solving $\frac{\partial \mathcal{L}}{\partial x_k^n} = 0$.
Denoting $ t = \frac{{x_k^n}^*}{{\alpha_k^n}^*}$,  $\frac{\partial \mathcal{L}}{\partial x_k^n} = 0$ is equivalent to the following quadratic equation.
\begin{equation}
(d^n_k+1)t^2 +(d^n_k+2)e^n_kt + (e^n_k)^2 - \frac{d^n_ke^n_k}{\lambda_k \ln 2}=0. \label{Qua_Eq}
\end{equation}
Note that the discriminant of this quadratic equation is $\Delta= {e^n_kd^n_k}^2 + \frac{4{d^n_k}^2}{\lambda_k\ln 2} + \frac{e^n_kd^n_k}{\lambda_k\ln 2}$, indicating that (\ref{Qua_Eq}) has two real roots.
According to the quadratic solution formula, we define $t^n_k(\lambda_k) = \frac{-(d^n_k+2)e^n_k + \sqrt{\Delta}}{2(d^n_k+1)}$.
Given that ${x^n_{k}}^* \in [0,\alpha_k^nQ^n_k]$, we can conclude that 
\begin{equation}
{x_k^n}^* = {\alpha_k^n}^* {T_k^n}^*, \label{Opt_xkn}
\end{equation}
where ${T_k^n}^*= [t^n_k(\lambda_k)]^{Q_k^n}_0$, and $[x]^a_b = \min\{\max\{x,b\},a\}$.
According to (\ref{div2}), it follows that
\begin{equation}
{\alpha_k^n}^*=\left\{\begin{array}{llll}
1, & \text{if } H_k^n > \beta_n \\ 
0, & \text{if } H_k^n < \beta_n
\end{array}\right. \label{Opt_bkn}
\end{equation}
where $H_k^n = {R^n_{k}}\left( {T_k^n}^*\right) - {T_k^n}^*{R^n_{k}}'\left({T_k^n}^*\right)$.
If $H_k^n$ are all different for $k \in \mathcal{K}$, according to constraint $(\ref{st_asg})$ and (\ref{Opt_bkn}), we have 
\vspace{-0.75em}
\begin{equation}
{\alpha_{k'}^n}^*=1, {\alpha_{k}^n}^*=0\label{Opt_akn}, \forall k \neq k',
\end{equation}
where $k' = \arg \underset{k}{\max} H^n_{k}$.
For SC $n$, only the D2D pair with the largest $H_k^n$ should be assigned this SC.
Note that the value of $\lambda_k$ can be determined by the sub-gradient method \cite{bertsimas2005optimization}.
The updating procedure of $\lambda_k$ in the $(t+1)$-th iteration is 
\begin{equation}
	\lambda_k^{(t+1) }=\left[ \lambda_k^{(t)}- \theta_{k}^{(t)}\left(P^D_{\max} - \sum_n^N (x_k^n)^{(t)}\right)\right]^{+}. \label{lambda}
\end{equation}
where $[a]^+ = \max\{0,a\}$, and $\theta_{k}^{(t)}$ is the positive step size. 
According to \cite[Proposition 6.3.1]{bertsimas2005optimization}, the sub-gradient method converges to the optimal solution to Problem $\mathcal{P}3$ for sufficient small step size $\theta_{k}^{(t)}$.
Thus, the transmit power of the D2D pairs can be obtained as $q_k^n = \alpha_k^n x_k^n$.
Overall, the above analysis is summarized as the following dual-based iterative resource allocation (DBIRA) algorithm to solve problem $\mathcal{P}1$. 

\section{Simulation Results}

The performance of the proposed resource allocation scheme is evaluated by simulations in this section.
The cell is a 500 m $\times$ 500 m square area with BS located in center.
The maximum distance between each D2D transmitter and receiver is 30 m.
The rate requirements for CUs are the same and denoted by $\gamma_{th}$.  
We set $N=30$, $P^C_{\max}=35$ dBm, $P^D_{\max}=25$ dBm, and $\sigma^2= -114$ dBm.
The Okumura-Hata loss model is adopted and the standard deviation of log-normal shadow fading is 4 dB.
All results are averaged over 1000 random realizations.
For comparison, we adopt the orthogonal frequency division multiple access (OFDMA) system that have multiple CUs on each SC as the benchmark, labeled as the MCU-OFDMA scheme, where the joint power control and channel assignment algorithm in \cite{SPL} is applied.
In MCU-OFDMA system, each SC is also shared by $M$ CUs, but each CU is only allowed to access $\frac{1}{M}$ fraction of SC bandwidth, so that the multiplexed D2D pair in MCU-OFDMA is also interfered by $M$ co-channel  CUs. 

\vspace{-1em}
\begin{algorithm}[H]
	\caption{ {Dual Based Iterative Resource Allocation (DBIRA) Algorithm}}
	\begin{algorithmic}\label{alg1}
		\STATE Initialize ${x_k^n}^{(0)}=0,{\alpha_k^n}^{(0)}=0, \forall k \in \mathcal{K}, n \in \mathcal{N}$.\\ 
		\STATE Initialize $\lambda_k^{(0)}$, step size $\theta_{k}^{(0)}$ for all $k \in \mathcal{K}$, and set the precision $\epsilon$.\\ 
		\REPEAT
		\FOR{$n \in \mathcal{N}$,$k \in \mathcal{K}$}
		\STATE  Calculate ${\alpha_k^n}^{(t)}$ and ${x_k^n}^{(t)}$ according to (\ref{Opt_akn}) and (\ref{Opt_xkn}), respectively;
		\ENDFOR
		\STATE Update $\lambda_k^{(t)}$ according to (\ref{lambda}) ;
		\STATE Update ${R_{\text{max}}^D}^{(t)}$ according to (\ref{P3obj}) ;
		\UNTIL  $|{R_{\text{max}}^D}^{(t)}- {R_{\text{max}}^D}^{(t-1)}|< \epsilon$ ;	
		\STATE Calculate $q_k^n$ and $p_k^n$ according to (\ref{pi}) for all $k \in \mathcal{K}, n \in \mathcal{N}$;
		\ENSURE $q_k^n$, $p_k^n$, ${\alpha_k^n}^{(t)}$, ${R_{\text{max}}^D}^{(t)}$.
	\end{algorithmic}
\end{algorithm}
\vspace{-1em}

\begin{figure}
\begin{minipage}[t]{0.5\textwidth}
\centering
\vspace{-2.0em} 
\includegraphics[width=1\textwidth]{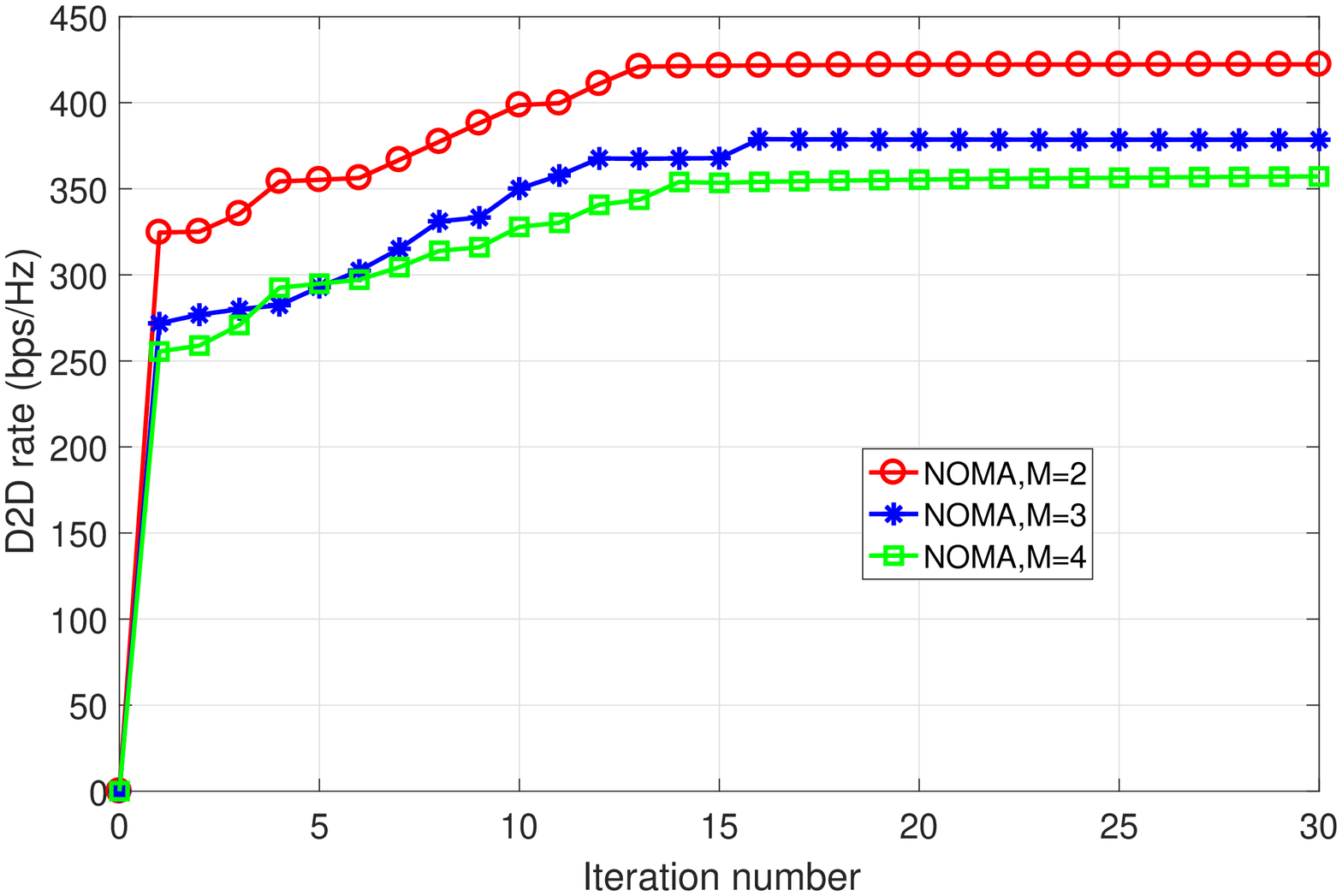}
\vspace{-2.5em}
\caption{ {Convergence performance of DBIRA algorithm.}}
\vspace{-2.0em}
\label{fig0}
\end{minipage}
\begin{minipage}[t]{0.5\textwidth}
\centering
\vspace{-2.0em} 
\includegraphics[width=1\textwidth]{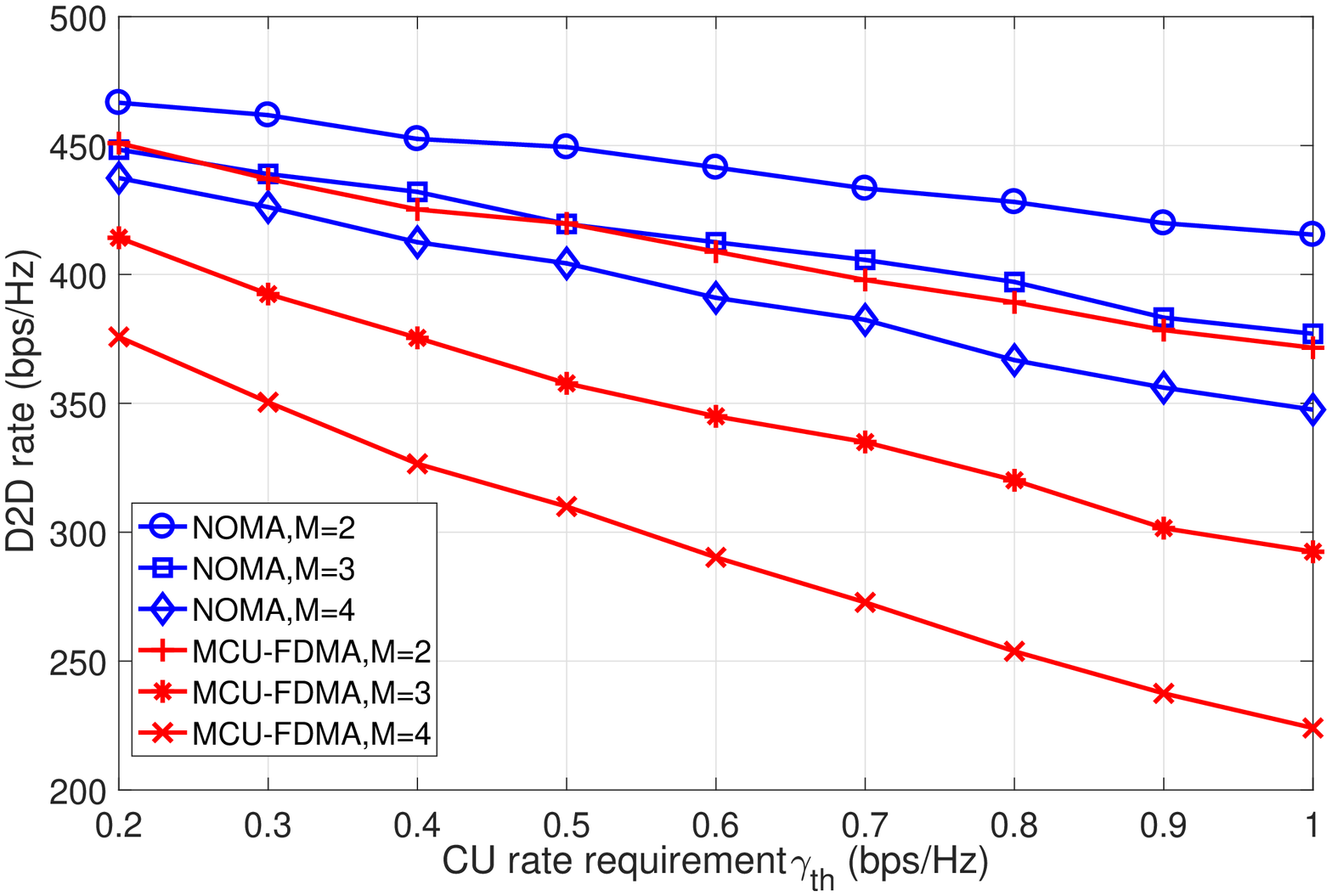}
\vspace{-2.5em}
\caption{ {D2D sum rate w.r.t CUs' rate requirements.}}
\vspace{-2.0em}
\label{fig1}
\end{minipage}
\end{figure}

Fig. \ref{fig0} illustrates the convergence behavior of the proposed DBIRA algorithm versus the number of iterations under different $M$.
As expected, it is shown that the sum rate of D2D pairs monotonically increases during the initial iterations.
Moreover, the sum rate performance converges within 20 iterations for all considered three cases, which validates the effectiveness of the proposed DBIRA algorithm.

Fig. \ref{fig1} shows the sum rate of D2D pairs w.r.t CUs' minimum rate requirements $\gamma_{th}$ under different $M$.
As expected, the NOMA-based scheme outperforms MCU-OFDMA scheme, especially when the number of CU multiplexed on each SC is large. 
The CUs need larger transmit power in MCU-OFDMA scheme to satisfy the same rate requirement, compared with NOMA-based scheme.
This leads to larger interference to the D2D pairs in MCU-OFDMA scheme than that in NOMA-based scheme, since the interferences to D2D pairs are summed from all multiplexed $M$ CUs in both MCU-OFDMA and NOMA schemes.
Moreover, the sum rate of D2D pairs decreases with the rate requirements of cellular links, which is also due to the larger transmit power for CUs required by the higher data rate requirements.

\section{Conclusion}
The resource allocation problem for D2D communications underlaying a NOMA-based cellular network was investigated in this letter.  
Although additional power constraints are introduced to D2D pairs for the sake of the NOMA decoding order, 
it is shown that the D2D underlaying NOMA cellular network still outperforms the conventional scheme for the network with high data requirements and myriad users.

\renewcommand{\baselinestretch}{1.40}
\bibliographystyle{IEEEtran}
\bibliography{IEEEabrv,Refer_03_27}

\end{document}